\documentclass[prl,preprint,superscriptaddress,amsmath,amssymb,floatfix]{revtex4}
\usepackage{graphicx,color}
\usepackage{dcolumn}
\usepackage{bm}
\usepackage{setspace}

\usepackage{color}

\usepackage{cancel}
\usepackage{cleveref}

\begin{document}

\title{A superconducting microwave multivibrator produced by coherent feedback}
\author{Joseph Kerckhoff} \email{jkerc@jila.colorado.edu}
\author{K. W. Lehnert}
\address{JILA, National Institute of Standards and Technology and the University of Colorado, Boulder, Colorado 80309, USA}

\date{\today}

\begin{abstract}
We investigate a nonlinear coherent feedback circuit constructed from pre-existing superconducting microwave devices.  The network exhibits emergent bistable and astable states, and we demonstrate its operation as a latch and the frequency locking of its oscillations.  While the network is tedious to model by hand, our observations agree quite well with the semiclassical dynamical model produced by a new software package [N. Tezak {\it et al.}, arXiv:1111.3081v1] that systematically interpreted an idealized schematic of the system as a quantum optic feedback network.
\end{abstract}

\maketitle


\noindent \textcolor{black}{The degree of control over matter and electromagnetic fields demonstrated in the past two decades suggests that quantum engineering may become a powerful discipline.  However, the extreme requirements for quantum scale engineering, be it for quantum--~\cite{MI} or ultra--low energy classical--information systems~\cite{Mill10}, suggest that active feedback will be necessary in useful networks \cite{MI, Jone10,Mabu11b}.  But while important proof--of--principle demonstrations of quantum error correction (QEC) have been reported for instance \cite{Reed12,Schi11,Aoki09}, the unwieldy \emph{classical} feedback equipment so far employed poses perhaps the greatest obstacle to realizing useful, complex systems.  More generally, measurement-based quantum feedback \cite{Smit02,Arme02,Sayr11,Vija12,Rist12} may prove impractical simply because ``measurement'' implies a network interruption by a fundamentally non-integrable system.  To overcome this bottleneck, quantum networks may need to actively stabilize themselves through coherent feedback of probes without measurement \cite{Reed12,Schi11,G&J,LQG,Hame12,Kerc10,Kerc11a,Mabu11,Mabu08,Iida11}.  Moreover, \cite{LQG,Hame12} suggests that coherent feedback can outperform even ideal measurement--based feedback.}  

\textcolor{black}{Measurement-based feedback to superconducting microwave quantum circuits is particularly difficult as signal transfer between a cryostat and room temperature electronics is inefficient and slow \cite{Vija12,Rist12,Sidd11}.  It was proposed in \cite{Mabu11} that coherent feedback circuits employing non--linear (Kerr) resonators are a natural approach to self--stabilizing, digital optical information processing in a complex quantum network.  Here, we demonstrate that these insights readily apply to superconducting circuits by constructing a coherent feedback multivibrator network (a circuit operable as a set--reset latch or an astable oscillator) from pre--existing Kerr-type resonators and coherent feedback of signals that never leave the $<$50mK environment.  This network becomes useful when integrated with other systems, and could act as a binary controller in a larger QEC coherent feedback network \cite{Kerc10,Kerc11a} or as a cryogenic clock.  And while an idealized model of this device could be derived manually, more complex systems would prove intractable.  Thus, we demonstrate that our observations agree quite well with a semiclassical model that was systematically produced from a network schematic by a hierarchical quantum circuit modeling package \cite{Teza11}.  While previous experiments have validated similar approaches to modeling coherent feedback circuits in linear \cite{Mabu08} and linear--quantum \cite{Iida11} optical networks, to our knowledge this is the first application to a nonlinear network, in a superconducting microwave context, and using automated quantum circuit modeling.}

The network's primary components are two single port microwave resonant circuits whose resonance frequency is power dependent and tunable with an applied magnetic flux \cite{Cast08}.  These tunable Kerr circuits (TKCs) were originally fabricated to serve as Josephson parametric amplifiers for near quantum--limited amplification of weak microwave signals and the preparation of squeezed microwave fields \cite{Mall11,Teuf11} (see also \cite{SMAmp}).  The TKCs are quarter--wave transmission line resonators formed by a coplanar waveguide with one end shorted and a capacitively coupled port at the other, and were mounted in separate sample boxes.  A series array of 40 Josephson junction SQUIDS interrupt the coplanar waveguide center conductor, providing a non--linearity that makes the devices' input--output (I/O) properties analogous to that of a high--quality, single--sided optical Kerr cavity (with Kerr coefficient $\chi<0$) \cite{Cast08}.  Thus, the reflected phase is a non--linear function of input power~\cite{Yurk06}.  This function can even be bistable for input drives that simultaneously are detuned below the TKCs' center frequency $\omega_0$ by at least the critical value $\omega_0-\omega_{p,c}=\Delta_c=\sqrt{3}\kappa$ and exceed the critical power $P_c=\hbar\omega_p\times4\kappa^2/(3\sqrt{3}|\chi|),$ where $\kappa$ is the field decay rate of the TKC \cite{Yurk06}.  The TKCs used here both have $\kappa/2\pi=15$ MHz and $P_c=-98\pm2$ dBm (uncertainty in the line calibration) when tuned such that $\omega_0/2\pi=6.408$ GHz.

\begin{figure}[b!]
\includegraphics[width=0.75\textwidth]{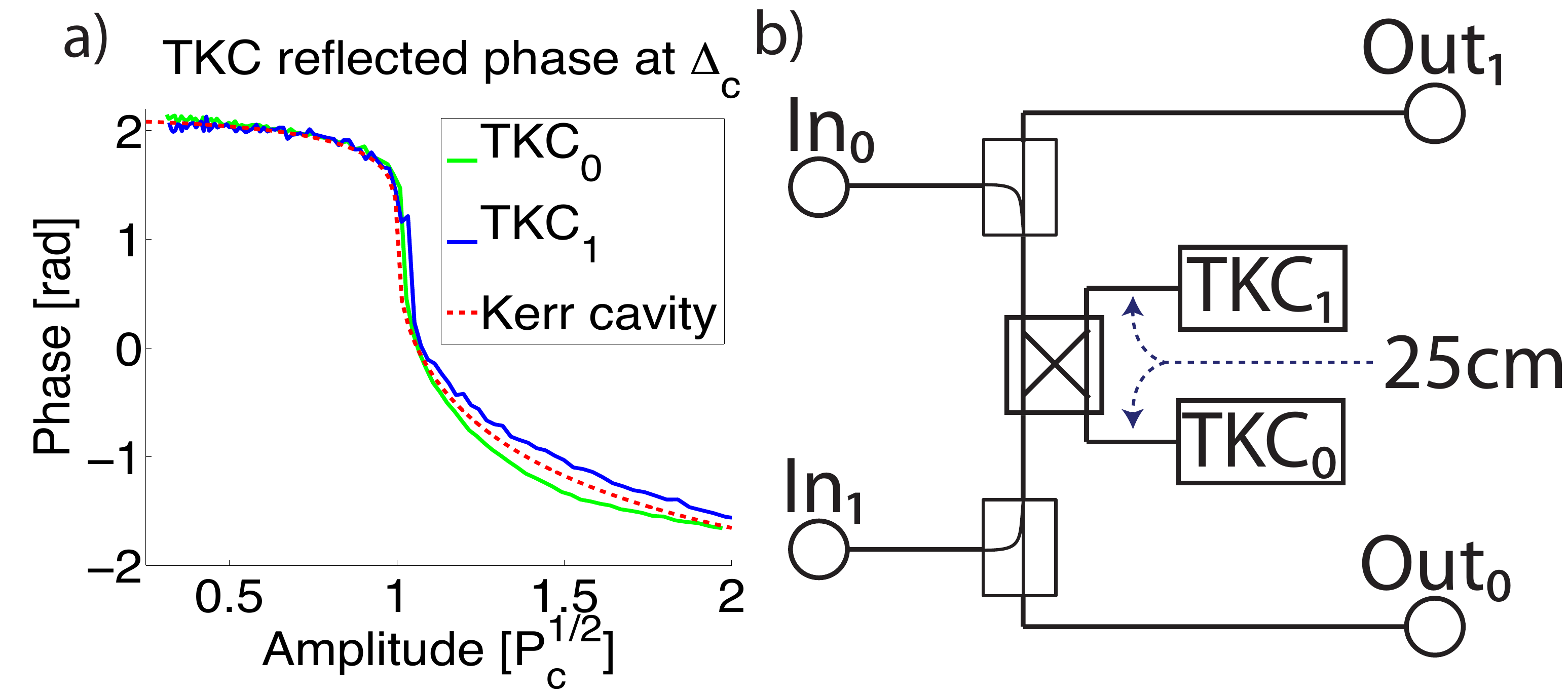}
\caption{\label{fig1} (color online) a) Each TKC operates as a single--sided Kerr cavity.  Reflected phase is a nonlinear function of drive amplitude for drive detunings $\lesssim\Delta_c$.  b) Network schematic.  Two flux biased TKCs are connected with approximately 25 cm cable connections to a quadrature hybrid in a feedback configuration.  Phase locked signals drive the ports In$_0$ and In$_1$, and are separated from output signals by directional couplers.}
\vspace{-0.1in}
\end{figure}

When the input drive detuning is close to, but does not exceed $\Delta_c$, a TKC is monostable for all input powers, but the phase of the reflected signal `flops' by approximately $\pi$ radians when the power, $p\times P_c$, exceeds $P_c$, see Fig.~\ref{fig1}a.  Because such phase shifts are readily converted into power variations in an interferometric network, \cite{Mabu11} suggested that Kerr cavities may usefully approximate NAND gates in an optical network.  For example, if two channels carrying either \textcolor{black}{$p\gtrsim1/4$} (high) or $p< 1/4$ (low) interfere in phase and are directed at a Kerr cavity, the phase of the reflected signal will `flop' only if both inputs are high.  Moreover, while a NAND gate/Kerr cavity is monostable in isolation, a network of two NAND gates/Kerr cavities in mutual feedback may function as a multivibrator.  Adapting \cite{Mabu11} to our context, a coherent feedback network of two TKCs should display emergent bistable and astable dynamics.  Such a network would also be nearly lossless and suitable for chip--level integration with quantum information microwave systems.

Represented in Fig.~\ref{fig1}b, the network components are housed in a dilution refrigerator and consist of two TKCs and a 4-8 GHz commercial quadrature hybrid (analogous to an optical 50/50 beamsplitter).  The TKCs are connected to the hybrid in such a way that signals they reflect are split between one of the two network outputs and the other TKC's input, producing a coherent feedback network.  These connections were made by low--loss, coaxial Cu cables, but our lab has previously interconnected these components on a single chip \cite{Ku11}.  Two signal generators drive the system through low--temperature attenuation stages, producing two phase--locked, low--temperature microwave drive inputs to the network.  The signals reflected out these same lines are separated from the inputs by directional couplers, and are amplified by two low--noise cryogenic HEMTs for analysis.       

\textcolor{black}{Superconducting microwave devices are often describable with models equivalent to I/O models in quantum optics \cite{Yurk84,RoMP}.  In such cases (e.g. TKCs and hybrids), one may model interconnected devices using cascaded I/O techniques still developing in quantum optics \cite{G&J,QN,Carm93}.  Unfortunately, these calculations are tedious, even for networks as basic as Fig.~\ref{fig1}b.  A new software package, Quantum Hardware Description Language (QHDL) \cite{Teza11}, adapts a standard electrical engineering modeling language to automate this modeling, interpreting a schematic diagram input that specifies the bosonic field I/O connections between pre--defined quantum optical primitive or composite models.}  

\textcolor{black}{Here, after a schematic representing Fig.~\ref{fig1}b is loaded, QHDL outputs the network's symbolic Heisenberg equations of motion (EOM).  As these TKCs typically contain $\sim$1000 photons, driven by coherent fields, a semiclassical approximation is invoked to simulate mean field dynamics, normal ordering the system operators in the EOM and replacing operators with complex scalars \cite{Supp}.  \textcolor{black}{While these approximations could have been applied at the device level, the quantum network model construction is no more difficult than the coherent classical one.  In our approach, future work considering the effects of intrinsic quantum fluctuations or integration with necessarily quantum systems follows readily.}  QHDL employs a standard approach to I/O theory that assumes that transmission line delays are negligible \cite{G&J,QN,Carm93}.  Furthermore, to compare our experiment to the ideal, in our model the TKCs are lossless and identical single--sided Kerr cavities, interconnections produce symmetric phase shifts and loss, and unwanted reflections are ignored.} 

\begin{figure}[b!]
\includegraphics[width=.75\textwidth]{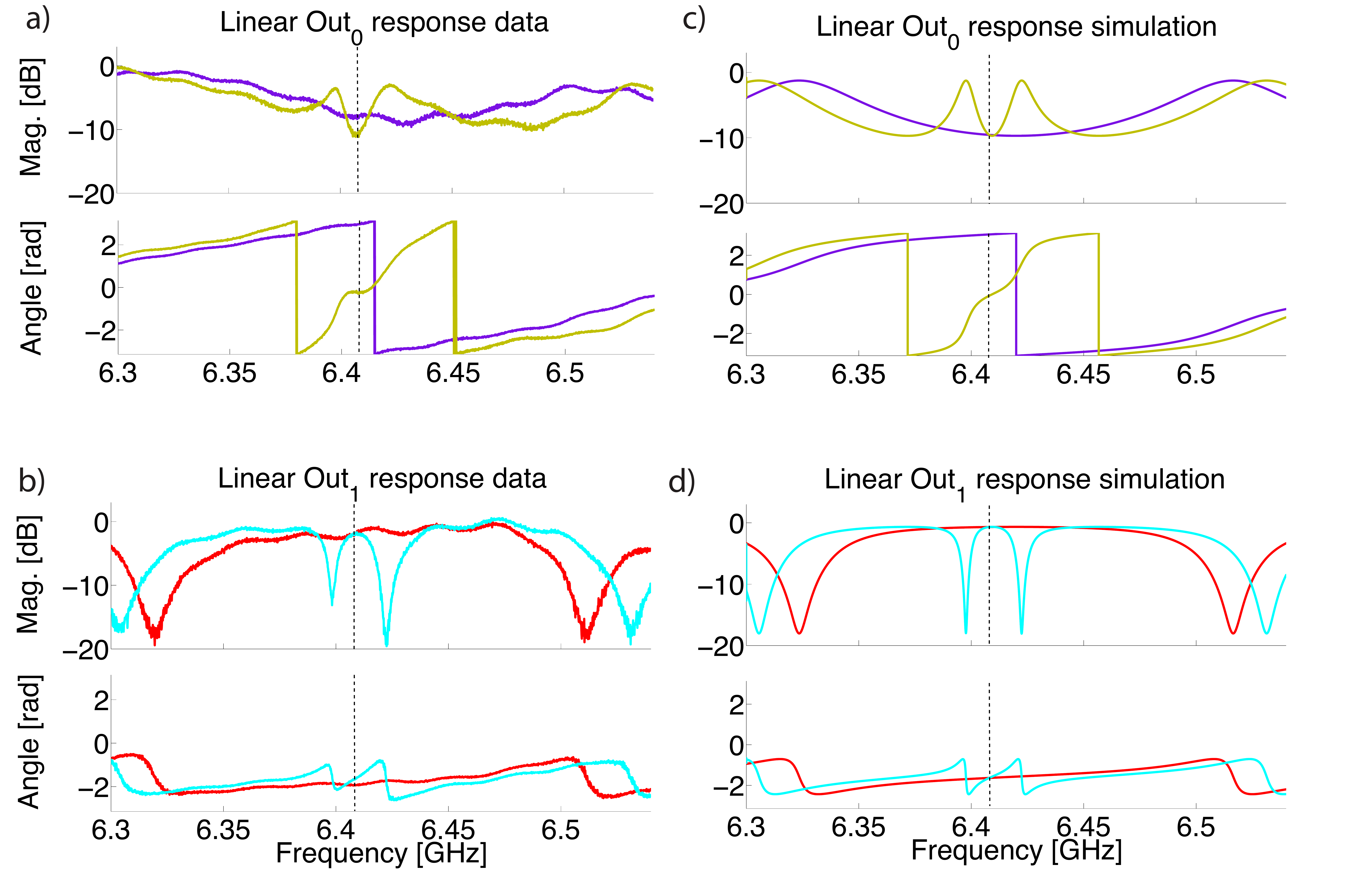}
\caption{\label{fig2} (color online) Magnitude and phase response data (a \& b) and simulation (c \& d) in the linear, $p\ll1$ regime.  Driving In$_1$ only and with the TKCs out of probe range, the response measured at Out$_0$ (Out$_1$) is shown in purple (red).  Gold (cyan) lines depict Out$_0$ (Out$_1$) responses when both TKCs are co--resonant at $\omega/2\pi=6.408$ GHz (dotted vertical lines).}
\vspace{-0.1in}
\end{figure}

In testing our model, we first measure the network's linear behavior by probing it with power $p\ll 1$. Tuning the TKCs far outside the probed region, we observe interferometric resonances, with the input power periodically distributed between the two outputs as the drive frequency varies.  With the TKCs tuned to be co--resonant near the middle of the probe range, additional resonances appear, and avoided crossings between all resonances are apparent, indicating that the TKCs are coherently coupled at rate $\approx\kappa$ to each other and to the network (Fig \ref{fig2}a-b).  Comparing the data with model simulations (Fig.~\ref{fig2}c-d), we calibrate 0.4 dB round--trip loss in each interconnection.  We note that the network's interconnections are longer than needed, a compromise between wanting long connections so that a desired phase shift between components could be achieved through frequency tuning and wanting short connections such that the delay between components be negligible.  Intending to consider dynamics only on time scales greater than $\kappa^{-1}$ when the experiment was deployed, we chose 25 cm interconnections (resulting in a .24$\kappa^{-1}$ delay between TKCs), producing a 385 MHz period in the frequency response.  

\begin{figure}[b!]
\includegraphics[width=0.75\textwidth]{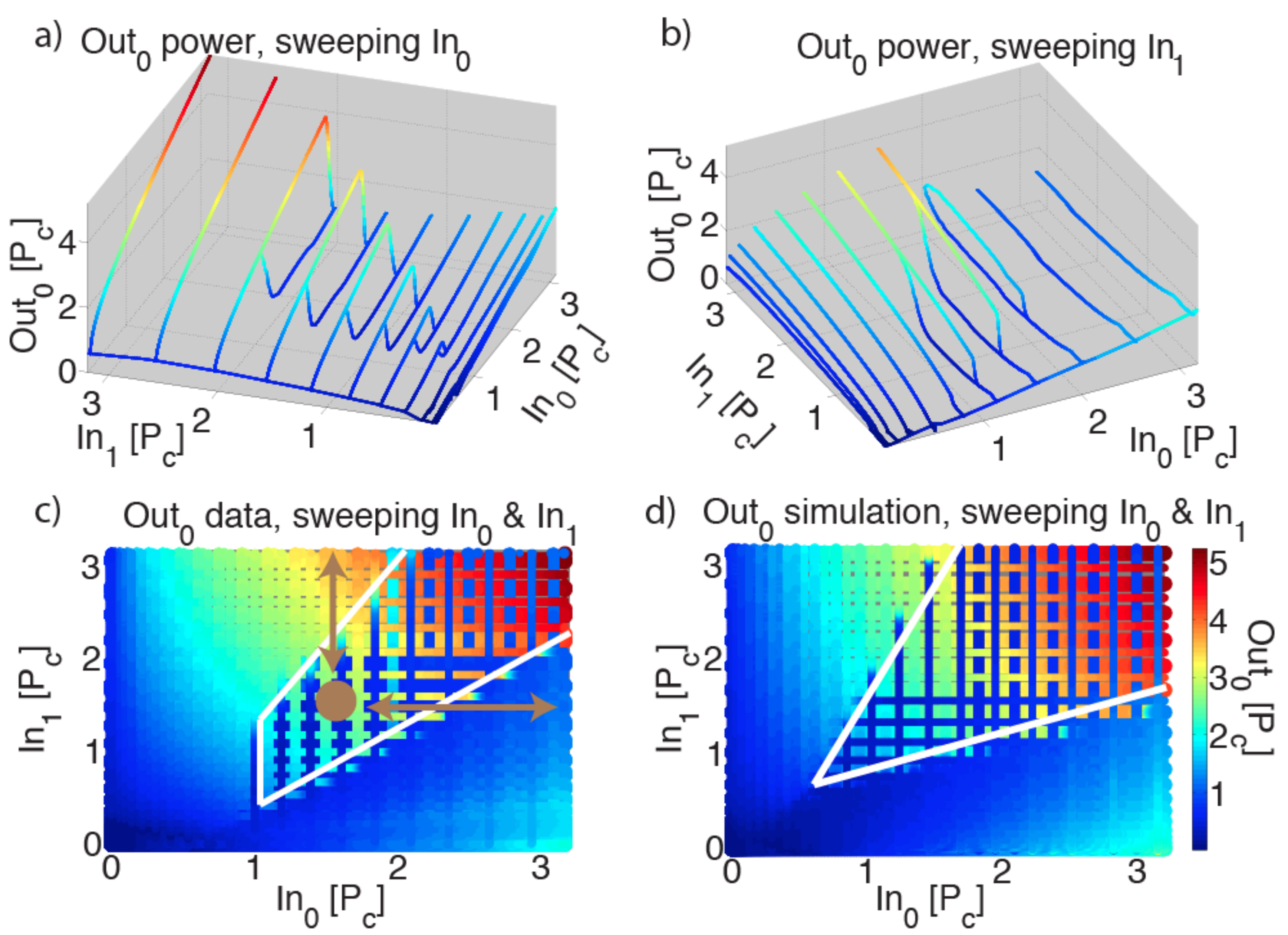}
\caption{\label{fig3} \textcolor{black}{(color online)  a-b) Mean Out$_0$ power, adiabatically sweeping either In$_0$'s or In$_1$'s drive amplitude high-and-low, with different biasings on the other input.  c)  Alternately sweeping in both directions produces a colored mesh (color indicating Out$_0$ power) where the bistable region is apparent where different colored lines intersect.  White lines demarcate the observed bistability region; brown dot represents the `hold' biasing state; arrows represent `set' or `reset'  operations.  d) Simulation of c.}}
\vspace{-0.1in}
\end{figure}

\textcolor{black}{Beyond the linear regime, and fixing both TKCs to $\omega_0/2\pi=6.408$ GHz and our probe frequency to $\omega_p/2\pi=6.39$ GHz (so that $\omega_0-\omega_p=0.69\Delta_c$ and the round trip phase shift per cable is 2.65 rad), the network exhibits both bistable and astable dynamics.  For balanced, $p\gtrsim 1$ and out--of--phase drives on the two inputs, coherent feedback causes the network to be bistable, with either high power driving TKC$_0$ and Out$_1$ and low power driving TKC$_1$ and Out$_0$, or vice versa.  To give a heuristic explanation (see \cite{Supp} for a quantitive model), if the power incident on TKC$_0$ happens to be high ($p\gtrsim1$), it is reflected with a `flopped' phase shift ($\approx\pi$ rad).  For these biasing conditions, the TKC$_0$-reflected signal interferes with the In$_1$ drive at the hybrid such that more power is directed to Out$_1$ than TKC$_1$.  The low power ($p<1$) signal incident on TKC$_1$ is then reflected with no additional phase shift, and consequently interferes at the hybrid with the In$_0$ drive such that more power drives TKC$_0$ than Out$_0$, reinforcing the original, strong TKC$_0$ drive.  By symmetry, for the same biasing conditions, the opposite network state is also self-stabilizing.  Thus, while both TKCs would be monostable in isolation at this detuning, the network exhibits a bistable output regime when the two input drives are balanced and strong.  If one further increases the In$_0$ (In$_1$) drive enough relative to the other, bistability disappears, and the system relaxes to a high Out$_1$ (Out$_0$) state.} 

\textcolor{black}{In Fig.~\ref{fig3}a-b, we plot the mean Out$_0$ power observed as a function of the input drives, as the amplitude of either the In$_0$ or In$_1$ drive is adiabatically swept high-and-low at 1 kHz for 100 cycles while the other input is fixed at various amplitudes.  Several hysteresis loops are apparent in the regime where the the two input drives are roughly equal, a consequence of the bistable dynamics described above.  The Out$_1$ powers are largely symmetric upon exchange of the input axes (asymmetries being a consequence of slight network asymmetries not considered here).  The bistable region as a function of the two inputs becomes clearer in Fig.~\ref{fig3}c, where low-to-high and high-to-low sweeps of one input are alternated for various static biasings of the other input, and the mean Out$_0$ power is depicted on the same color scale.  This produces a colored mesh that indicates the bistable region by the intersection of different color lines.  Fig.~\ref{fig3}d is a simulation of the same, producing a very similar color pattern and a similar, but more symmetric bistability region.  All output power data was calibrated first by scaling the signal measured at Out$_1$ such that for far--detuned TKCs and balanced inputs (inferred by the 6.3-6.55 GHz phase response) the output powers were equal (compensating for amplifier asymmetries), then by equally scaling both outputs such that the highest Out$_0$ power in Fig.~\ref{fig3}c matched the highest simulated power in Fig.~\ref{fig3}d.}

\begin{figure}[b!]
\includegraphics[width=0.75\textwidth]{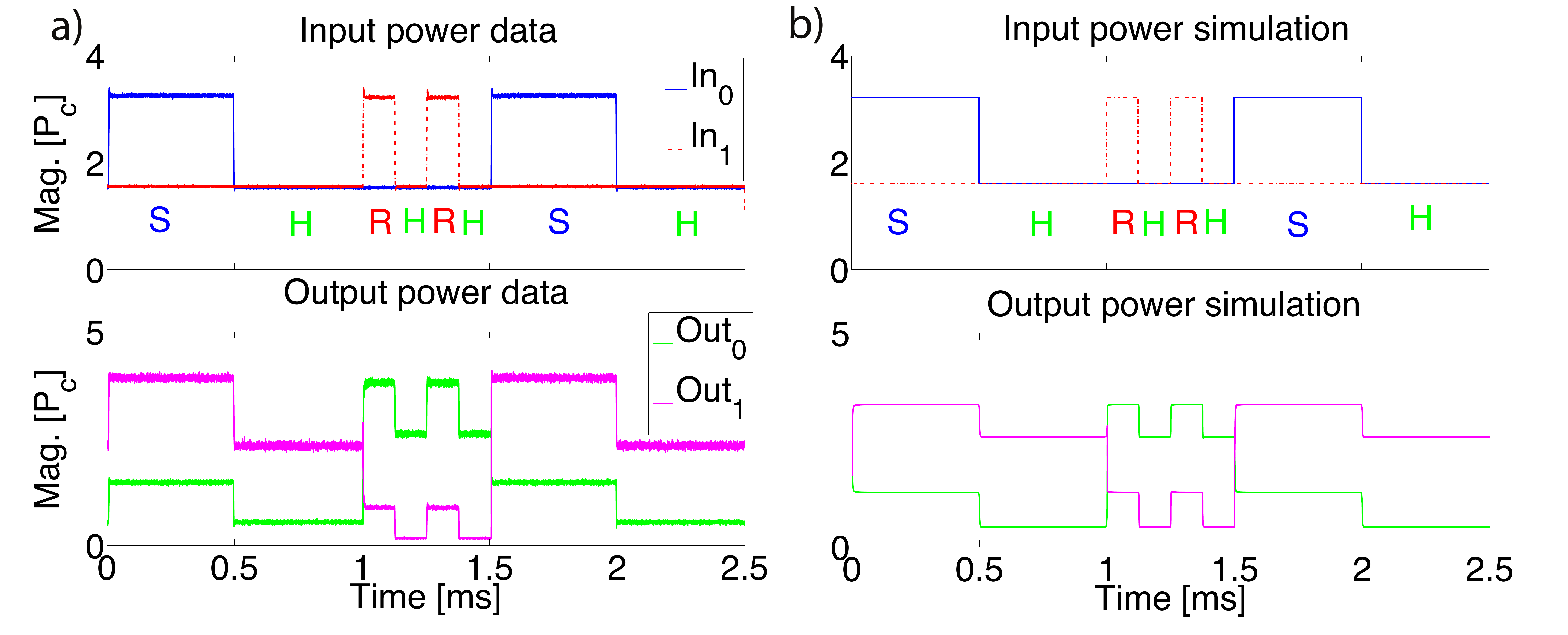}
\caption{\label{fig4}(color online) Mean output response data (a) and simulation (b) depicting the network's operation as a binary memory element.  After each `set'/`reset' operation (blue S/red R), input power is primarily directed out Out$_1$/Out$_0$, even after retuning to the `hold' state (green H).  Input states of variable durations were an experimental convenience.}
\vspace{-0.1in}
\end{figure}

 This bistability may be leveraged to operate the network as a set--reset latch (or `flip--flop'), a binary memory element that outputs power according to prior inputs \cite{Mabu11}.  In Fig.~\ref{fig4}a, the averaged output response is tracked as the two input drives are amplitude modulated between a `hold' condition of equal, $p=1.6$ drives, and either the In$_0$ (`set') or In$_1$ (`reset') drives doubling in power and returning.  Fig.~\ref{fig4}b simulates the same.  \textcolor{black}{The hold condition corresponds to the brown dot in Fig.~\ref{fig3}c, while the set and reset operations correspond to modulating the input powers according to the horizontal and vertical arrows, respectively.}  As the hold state is bistable and connected to the monostable set and reset states via different stable manifolds (Fig.~\ref{fig3}), each set--hold (reset--hold) event causes the Out$_1$ (Out$_0$) signal to swing high regardless of the prior state.  While the modulation frequency is again $\sim$kHz, the network's response rate is at least that of the 2 MHz detection bandwidth.  \textcolor{black}{To note one potential application, \cite{Kerc10,Kerc11a} suggests that set--reset sub--networks like these could act as binary controllers in `hard--wired' implementations of QEC, stabilizing superconducting qubit arrays in a larger coherent feedback network.}

\begin{figure}[b!]
\includegraphics[width=0.75\textwidth]{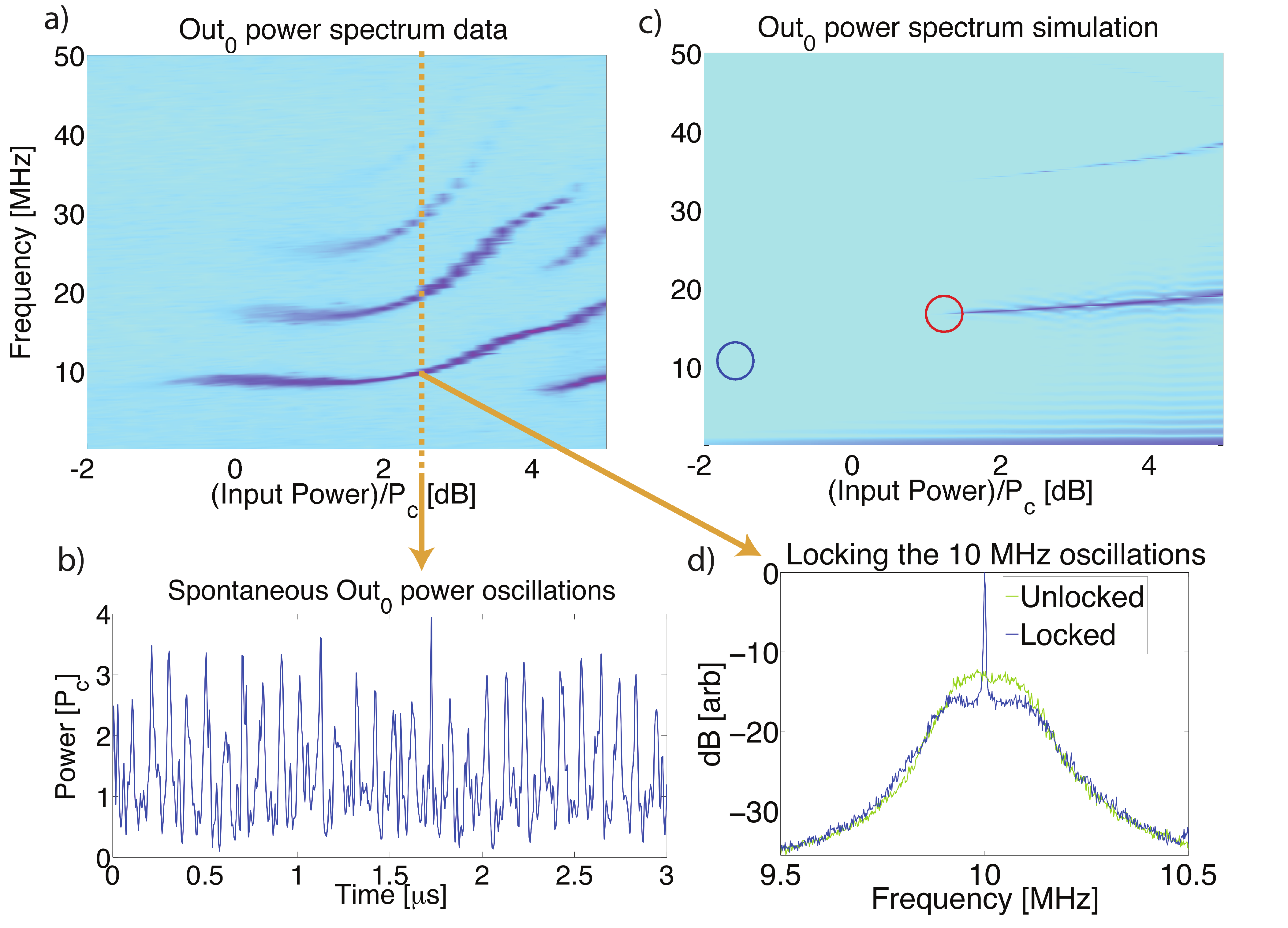}
\caption{\label{fig5} (color online) a) Mixed--down power spectrum detected at Out$_0$ as In$_0$ is driven at 6.39 GHz, input powers relative to $P_c$.  b) Out$_0$ power oscillations in time with 2.58 dB input.  c) Simulation of the Out$_0$ power spectrum.  Red (blue) circle marks the predicted emergence of a stable limit cycle when delays are not (are) added to a linearized model \cite{Supp}.  d) Power spectrum of the Out$_0$ signal frequency component 10 MHz detuned from the injected tone with and without frequency locking.}
\vspace{-0.1in}
\end{figure}

Increasing the detection bandwidth to 50 MHz, various drive settings produce sustained output power oscillations at frequencies $\approx\kappa$.  For example, Fig.~\ref{fig5}a represents the mixed--down power spectrum detected at Out$_0$ while driving only In$_0$ with a continuous wave 6.39 GHz tone of various amplitudes.  Starting near $p=-1$ dB, $\sim$10 MHz and higher harmonics emerge and accelerate with input power.  \textcolor{black}{To compare to the bistable case, with only one drive, a strong or weak signal reflected by TKC$_0$ has no drive to interfere with.  Thus, as TKC$_0$ equilibrates, TKC$_1$ is driven with a relatively strong or weak signal, respectively, the opposite of the bistable case.  Consequently, when this signal ultimately reflects back towards TKC$_0$, the network destabilizes and oscillates between both states \cite{Supp}.}  

In this case, however, the analogous simulation (Fig.~\ref{fig5}c) predicts 17 MHz power oscillations, their emergence at $p=1.2$ dB, and their increasing frequency with drive power.  In view of the accuracy of the idealized simulations to reproduce the low--frequency dynamics in Figs.~\ref{fig1}-\ref{fig4} and the significant frequency-dependence of phase shifts of dynamic signals $\gtrsim$10 MHz in our physically-extended network (see Fig.~\ref{fig2}), we suspect the discrepancy stems from the zero delay assumption of QHDL.  This hypothesis is supported by a linearized version of the QHDL-derived model.  Analysis of the dynamics about the EOM fixed points predicts the emergence of a stable 17 MHz limit cycle at 1.2 dB, exactly as observed in the Fig.~\ref{fig5}c simulation.  Adding the approximate effects of transmission line delays to the linearized model destabilizes the dynamics at lower drive powers, suggesting a stable 11 MHz limit cycle emerging at -1.6 dB \cite{Supp}, much closer to what is experimentally observed (Fig.~\ref{fig5}a).   \textcolor{black}{I/O models may be generalized to include finite delays and while the resulting models may be automated, they were deemed too complex for first generation software \cite{G&J,Teza11}.}  Cascaded I/O models are most appropriate for chip--scale systems as opposed to our extended network; chip--scale integration would improve simulation accuracy if our hypothesis is correct.  \textcolor{black}{It is worth mentioning, though, that QHDL's qualitative accuracy beyond its range of strict applicability was quite useful for predicting astable parameter regions.}

Finally, we demonstrate (measurement--based) stabilization of these oscillations in Fig.~\ref{fig5}d.  By setting the In$_0$ drive to $p=2.58$ dB and mixing down the Out$_1$ signal with a 6.4 GHz local oscillator (10 MHz detuned from the injected tone) significant phase noise relative to our room temperature frequency standard is apparent (likely due to technical jitter in the TKC center frequencies).  As the frequency of the output power oscillations varies with input power, using this phase signal to drive .23 dB analog amplitude modulation of the injected tone (100 kHz modulation bandwidth) creates a phase locked loop that stabilizes the 10 MHz pulse train spontaneously produced by our cryogenic network to the 10 MHz room temperature clock that phase locks our generators.    

While these dynamics are classical, QHDL outputs quantum models and TKCs are routinely used by our lab to generate and measure non--trivial quantum fields \cite{Mall11}.  It would be interesting, for instance, to consider how quantum field fluctuations propagate through this network and perhaps disturb the mean--field dynamics reported here \cite{Kerc11}.  Nonetheless, classical dynamics are sufficient to demonstrate that classical information systems are readily produced by coherent feedback on generic quantum devices.  But because they are constructed from the same hardware as quantum microwave circuits, they hold a natural advantage in terms of the chip--level classical/quantum integration that would be necessary for truly scalable quantum circuits \cite{Kerc10,Kerc11a}.  \textcolor{black}{We conclude by reiterating that this system was constructed from pre--existing components of types generically available in superconducting circuit labs \cite{Cast08,SMAmp}.  And while this system's intricate and potentially useful dynamics are difficult to consider manually, they are readily analyzed and integrated into larger network models using a laptop and a small number of I/O laws originally formulated for quantum optics.  This observation suggests that automated modeling techniques like QHDL are now needed to properly compliment quantum hardware advances.}   

\begin{acknowledgments}
We acknowledge partial support from the DARPA QuEST program and from the NSF Physics Frontier Center.  JK acknowledges the NRC for financial support, W. Kindle and H.-S. Ku for experimental advice, and N. Tezak and H. Mabuchi for very helpful discussions and the beta version of QHDL.
\end{acknowledgments}


\section{Supplementary Information}

\subsection{General modeling}
The power of the Quantum Hardware Description Language (QHDL) \cite{Teza11} modeling approach (which automates the quantum circuit ÔalgebraÕ of Gough
and James \cite{G&J}, which in turn generalizes earlier work on cascaded open quantum systems
by Carmichael \cite{Carm93} and Gardiner \cite{Gard93, QN}) stems from the fact that individual open quantum optical components are given the same succinct representation as interconnected networks of quantum optical  components.  

This representation consists of a triple $(\mathbf{S},\mathbf{L},H)$.  To describe briefly, $\mathbf{S}$ is an operator-valued, square {\it scattering matrix} that specifies how input (uni-directional), freely-propagating bosonic fields are directly scattered to output fields (as many vector indices as there are input fields), as in the action of a beamsplitter or microwave hybrid.  $\mathbf{L}$ is an operator-valued {\it coupling vector} that specifies how each field mode couples to the internal quantum degrees of freedom (if any) of component devices.  And $H$ is the {\it effective Hamiltonian} that specifies the internal dynamics of the devices, independent of the effects of the free fields.  

\begin{figure}[b!]
\includegraphics[width=0.75\textwidth]{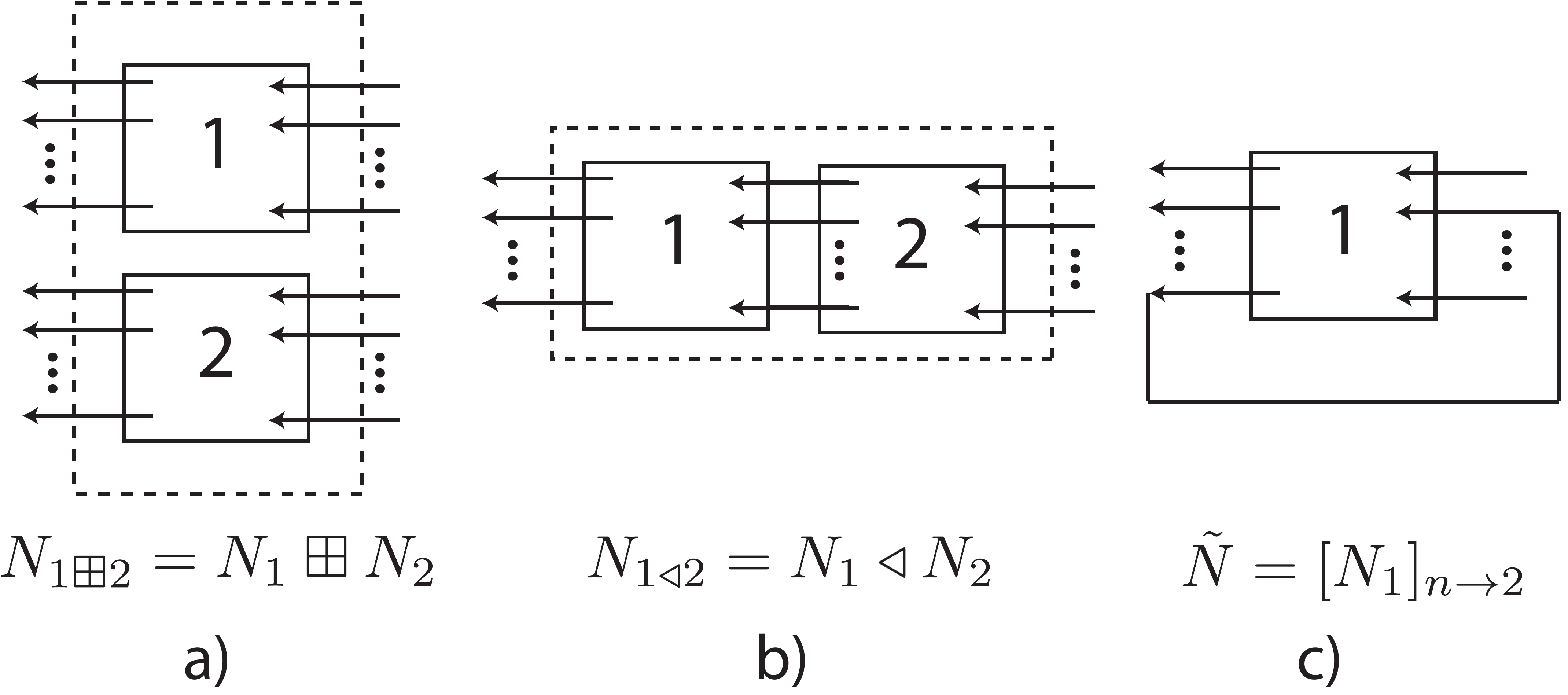}
\caption{\label{fig1_supp} Depictions of the essential composition operations through which component representations are combined to form composite network representations.  a) The concatenation product.  b) The series product.  c)  The feedback operation.  Adapted from \cite{Teza11}}
\vspace{-0.1in}
\end{figure}

The construction of a network $(\mathbf{S},\mathbf{L},H)$ from component triples proceeds with a small set of composition rules (here presented assuming negligible time delay between components), depicted in Fig.~\ref{fig1_supp}.  The {\it concatenation product} represents the effective dynamics of two components that have no direct free field interconnection, but could share a common internal Hilbert space:
\begin{equation}\label{eq:concatenation}
(\mathbf{S}_{1\boxplus2},\mathbf{L}_{1\boxplus2},H_{1\boxplus2}) = (\mathbf{S}_1,\mathbf{L}_1,H_1)\boxplus(\mathbf{S}_2,\mathbf{L}_2,H_2) = \left(\left[\begin{array}{cc} S_1 & 0\\0&S_2\end{array}\right],\left[\begin{array}{c}L_1\\L_2\end{array}\right],H_1+H_2\right).
\end{equation}
The {\it series product} represents the effective dynamics of a network in which the output fields of component 2 are fed into the inputs of component 1
\begin{equation}\label{eq:series}
(\mathbf{S}_{1\triangleleft2},\mathbf{L}_{1\triangleleft2},H_{1\triangleleft2}) = (\mathbf{S}_1,\mathbf{L}_1,H_1)\triangleleft(\mathbf{S}_2,\mathbf{L}_2,H_2) = \left(\mathbf{S}_1\mathbf{S}_2,\mathbf{L}_1+\mathbf{S}_1\mathbf{L}_2,H_2+H_1+\Im\{\mathbf{L}_1^\dag\mathbf{S}_1\mathbf{L}_2\}\right)
\end{equation}
 where $\Im\{A\} \equiv(A-A^\dag)/2i$ and the $^\dag$ operation returns a transposed operator matrix with operator adjoints in its entries.  Finally, the {\it feedback operation} represents the effective network dynamics when the $k^{th}$ output channel is fed back into the $l^{th}$ input (thus reducing the number of input and output ports by 1): $[(\mathbf{S},\mathbf{L},H)]_{k\rightarrow l} = (\tilde{\mathbf{S}},\tilde{\mathbf{L}},\tilde{H},)$ where
 \begin{eqnarray}
 \tilde{\mathbf{S}} &=& \mathbf{S}_{\cancel{[k,l]}}+\left[\begin{array}{c}S_{1,l}\\\vdots\\ S_{k-1,l}\\S_{k+1,l}\\\vdots\\S_{n,l}\end{array}\right](1-S_{k,l})^{-1}\left[\begin{array}{cccccc}S_{k,1}&\hdots& S_{k,l-1}&S_{k,l+1}&\hdots&S_{k,l}\end{array}\right]\nonumber\\
 \tilde{\mathbf{L}} &=& \mathbf{L}_{\cancel{[k]}}+\left[\begin{array}{c}S_{1,l}\\\vdots\\ S_{k-1,l}\\S_{k+1,l}\\\vdots\\S_{n,l}\end{array}\right](1-S_{k,l})^{-1}L_k\nonumber\\
 \tilde{H} &=& H + \Im\left\{\left[\sum_{j=1}^nL_j^\dag S_{jl}\right](1-S_{k,l})^{-1}L_k\right\}\label{eq:FB}
 \end{eqnarray}    
where $\mathbf{S}_{\cancel{[k,l]}}$ and $\mathbf{L}_{\cancel{[k]}}$ indicate the original scattering matrix and coupling vector with the $k^{th}$ row and $l^{th}$ column removed.  For more details of the fundamental models and assumptions, we refer readers to \cite{Teza11,G&J}. 

Whether a $(\mathbf{S},\mathbf{L},H)$ triple describes an individual component or a network of components, the effective dynamics of the system are calculated in the same way.  For example, assuming the input fields are in the vacuum state, the evolution of an operator $X$ that acts on the internal Hilbert space (e.g. the annihilation operator of a TKC mode) is systematically calculated as \cite{G&J}  ($\hbar=1$)
\begin{equation}\label{eq:EoM}
d X = \left(-i[X,H]+\frac12\mathbf{L}^\dag[X,\mathbf{L}]+\frac12[\mathbf{L}^\dag,X]\mathbf{L}\right)dt+d\mathbf{A}^\dag(t)\mathbf{S}^\dag[X,\mathbf{L}]+[\mathbf{L}^\dag,X]\mathbf{S}d\mathbf{A}(t)+\text{Tr}\left[(\mathbf{S}^\dag X\mathbf{S}-X)d\mathbf{\Lambda}^T(t)\right]
\end{equation} 
where $^T$ is the operator matrix transpose.  $\mathbf{A}(t)$, $\mathbf{A}^\dag(t)$ are operator vectors, whose entries are known as as {\it quantum noise processes}, whose infinitesimal increments (e.g. $dA_{[k]}(t)$) may be roughly considered the annihilation and creation operators (respectively) on the infinitesimal segment of input free field that interacts with the component or network at time $t$.  $\mathbf{\Lambda}$ is an operator matrix whose entries are a third kind of quantum noise process whose increments may be roughly considered bilinear products of field annihilation and creation operators (e.g. its diagonal elements are similar to number operators on each infinitesimal field segment).  Also, the output fields are related to the input fields and the internal degrees of freedom by
\begin{equation}\label{eq:IO}
d\mathbf{A}_{out}(t) = \mathbf{S}d\mathbf{A}(t)+\mathbf{L}dt,
\end{equation}
as well as related relations for $\mathbf{A}^\dag(t)$ and $\mathbf{\Lambda}(t)$.

Thus, when the assumptions are valid, the dynamics of both individual quantum optical components and complex networks of interconnected components may be derived systematically: following a schematic of interconnected $(\mathbf{S},\mathbf{L},H)$ models, one first derives the effective $(\mathbf{S},\mathbf{L},H)$ for the entire network using rules Eqs.~(\ref{eq:concatenation}-\ref{eq:FB}); then, one derives the quantum equations of motion using Eqs.~(\ref{eq:EoM}-\ref{eq:IO}).  Often, however, this general procedure is very tedious.  

The most immediate value of the Quantum Hardware Description Language (QHDL) \cite{Teza11} is that it insulates a user from this computational tedium.  {\it One may produce the desired equations of motion from an intuitive schematic diagram and less than 10 lines of code.}

\subsection{Specific model}
\begin{figure}[b!]
\includegraphics[width=0.45\textwidth]{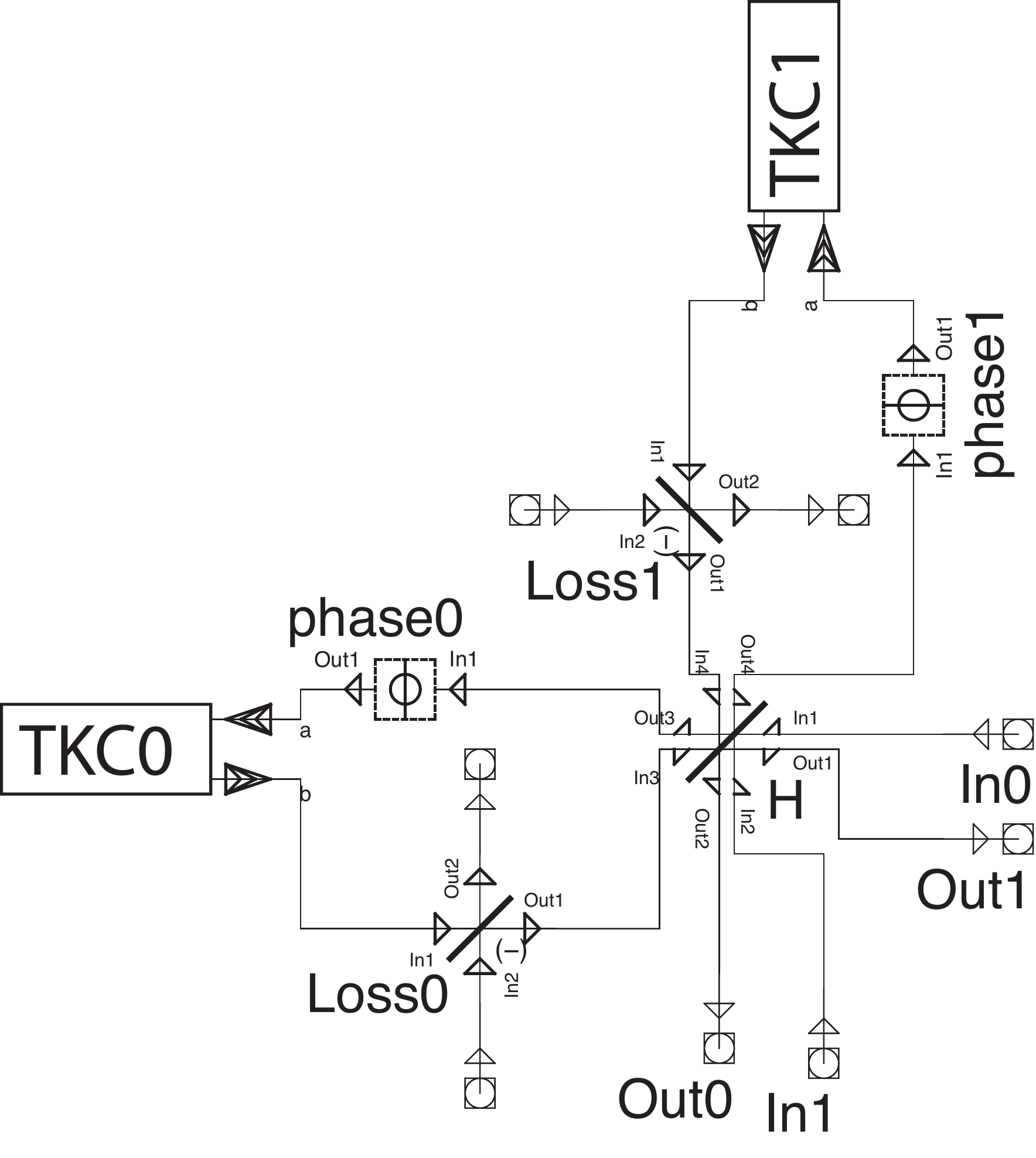}
\caption{\label{fig2_supp}  Schematic representation of the network model employed in the main Letter and interpreted by QHDL.  Overall a 4-mode input-output network (2 signal channels, 2 loss channels), individual components are icons representing quantum optical $(\mathbf{S},\mathbf{L},H)$ models, with connections between components representing (uni-directional) bosonic field modes.  Coherent drive `components' are not shown, but are eventually placed upstream of In$_0$ and In$_1$ ports.  This schematic and the sub-componets it references define the network model returned by QHDL.}
\vspace{-0.1in}
\end{figure}

Following this general modeling and using procedures analogous to \cite{Yurk06,RoMP,Yurk84}, one may derive the $T\equiv(\mathbf{S}_{TKC},\mathbf{L}_{TKC},H_{TKC})$ triple representation for an ideal TKC as a single mode component:
\begin{eqnarray}\label{eq:TKC}
\mathbf{S}_{TKC} &=& [-1]\nonumber\\
\mathbf{L}_{TKC} &=& [-i\sqrt{2\kappa}a]\nonumber\\
H_{TKC} &=& \Delta a^\dag a+\frac\chi2 a^{\dag2}a^2
\end{eqnarray}
where $a$ is the annihilation operator on the TKC resonator mode, $\Delta=\omega_0-\omega_p$ is the detuning between the TKC resonance frequency ($\omega_0$) and the carrier frequency of the input field driving the TKC ($\omega_p$), $\kappa$ is the field decay rate, and $\chi<0$ is the effective Kerr coefficient produced by the SQUID array.  The remaining component types employed in the network model are: beamsplitters $BS\equiv(\mathbf{S}_{BS},\mathbf{L}_{BS},H_{BS})$
\begin{equation}\label{eq:BS}
\mathbf{S}_{BS} = \left[\begin{array}{cc}\mu & -\nu^\ast\\\nu & \mu\end{array}\right],\quad \mathbf{L}_{BS} = \left[\begin{array}{c}0\\0\end{array}\right],\quad H_{BS}= 0
\end{equation}
where $|\mu|^2+|\nu|^2=1$; phase shifters $\Phi\equiv(\mathbf{S}_{\phi},\mathbf{L}_{\phi},H_{\phi})$
\begin{equation}\label{eq:phase}
\mathbf{S}_{\phi} = [e^{i\phi}],\quad \mathbf{L}_{\phi} = [0],\quad H_{\phi} =0;
\end{equation}
and coherent drives $W_\alpha\equiv(\mathbf{S}_{W\alpha},\mathbf{L}_{W\alpha},H_{W\alpha})$
\begin{equation}\label{eq:W}
\mathbf{S}_{W\alpha} = [1],\quad \mathbf{L}_{W\alpha} = [\alpha],\quad H_{W\alpha} =0,
\end{equation}
with complex amplitude $\alpha$.  From these general component models, the TKCs are taken as distinct but identical $T$ components.  The quadrature hybrid is modeled as the concatenation of two beamsplitters, $H \sim BS_0\boxplus BS_1$, with appropriate relations between the reflection and transmission coefficients $\{\mu_0,\nu_0\}$ and $\{\mu_1,\nu_1\}$ stemming from the fact that this single, bi-directional physical component is modeled as two uni-directional beamsplitters (with the ``$\sim$'' representing the fact that some field index re-ordering is also employed).  Transmission line-induced phase shifts are modeled as two identical $\Phi$ components, and transmission line loss is modeled by two identical beamsplitters that mix the transmission line modes with vacuum at a low rate, i.e. $|\nu|\ll|\mu|$.  The two coherent drives are modeled as two $W_\alpha$ components `upstream' of the network, which displace input vacuum fields by respective amplitudes.

Icons that represent these components are arranged in the schematic diagram shown in Fig.~\ref{fig2_supp}, with interconnections that emulate our experimental network.  From this schematic, QHDL parsers were employed to calculate first the effective $(\mathbf{S},\mathbf{L},H)$ representation of the network and then semiclassical approximations of its equations of motion.  To give a concrete example of the calculation procedure, we will devote most of the remainder of this section to outlining the procedure and results obtained in the case of an even simpler network model that is lossless and has integer $\pi$-radian phase shifts. 

If one removes the `Loss0' and `Loss1' beamsplitter components and associated input and output ports in the Fig.~\ref{fig2_supp} schematic, the network model without any coherent drives may be characterized as
\begin{equation}
N_{vac} = P_{(1,0)}\triangleleft\left[\left(I_2\boxplus(T_0\triangleleft\Phi_0)\right)\triangleleft\left[\left(I_3\boxplus(T_1\triangleleft\Phi_1)\right)\triangleleft H\right]_{4\rightarrow4}\right]_{3\rightarrow3}
\end{equation}
where we have introduced two new types of $(\mathbf{S},\mathbf{L},H)$ `components' necessary for appropriate field indexing: the {\it permutation matrix} $P_{(1,0)}$ that reverses the ordering of the two output fields and the {\it identity component} $I_n$ that passes $n$-input modes to outputs without scaling or re-ordering.  In plain English this sequence may be read as 
\begin{quote}
``Output 4 of $H$ is fed into $\Phi_1$ is fed into $T_1$ is fed back into input 4 of $H$.  Output 3 of $H$ is fed into $\Phi_0$ is fed into $T_0$ is fed back into input 3 of $H$.  The remaining two outputs are reordered.''
\end{quote} 
To represent the coherently driven dynamics, one then calculates
\begin{equation}
N = N_{vac}\triangleleft(W_{\alpha_0}\boxplus W_{\alpha_1}).
\end{equation}
If one then plugs in a quadrature hybrid model for $H$ (i.e. $\mu=1/\sqrt{2}$, $\nu=i/\sqrt{2}$) and sets both phase shifts to $\pi$, the resulting symbolic $N\equiv(\mathbf{S}_{N},\mathbf{L}_{N},H_{N})$ triple is relatively simple
\begin{eqnarray}
S_N &=& \left[\begin{array}{cc} 2\sqrt{2}i/3 & 1/3\\1/3 & 2\sqrt{2}i/3\end{array}\right]\nonumber\\
L_N &=& \left[\begin{array}{c} \frac{-\sqrt{2\kappa}}{3}a_0+2i\frac{\sqrt{\kappa}}{3}a_1+2i\frac{\sqrt{2}}{3}\alpha_0-\frac13\alpha_1\\
 \frac{-\sqrt{2\kappa}}{3}a_1+2i\frac{\sqrt{\kappa}}{3}a_0-2i\frac{\sqrt{2}}{3}\alpha_1+\frac13\alpha_0
 \end{array}\right]\nonumber\\
H_N &=& \Delta_0a_0^\dag a_0 + \Delta_1 a_1^\dag a_1+\frac{\chi_0}{2} a_0^{\dag2}a_0^2+\frac{\chi_1}{2} a_1^{\dag2}a_1^2+\left(-\frac{\sqrt{\kappa}}{3}a^\ast_0\alpha_0 +i \frac{\sqrt{2\kappa}}{6}a^\ast_0\alpha_1-i\frac{\sqrt{2\kappa}}{6}a^\ast_1\alpha_0 + \frac{\sqrt{\kappa}}{3}a^\ast_1\alpha_1+h.c.\right)
\end{eqnarray}
where $a_{\{0,1\}}$ is the annihilation operator for $T_{\{0,1\}}$, analogous labeling applies to $\Delta_i$ and $\chi_i$, and $\alpha_{\{0,1\}}$ are the coherent drive amplitudes driving inputs 0 and 1.

At this stage, one could produce the the full quantum mechanical equations of motion.  However, we also invoke a semiclassical approximation that is appropriate for our measurements in the main Letter.  That is, we instead calculate the equations of motion for the expectations of the degrees of freedom (e.g. $\tilde{a}_i\equiv\langle a_i\rangle$) and assume that the expectations of normal-ordered operators factor (e.g. $ \langle a_i^\dag a_i\rangle \approx |\tilde{a}_i|^2$).  Moreover, as the inputs to $N$ are vacuum fields (recall, the coherent drives that excite the network are actually part of $N$), all the noise terms drop out of these expressions and using Eqs. (\ref{eq:EoM}-\ref{eq:IO}), we are left with a closed system of equations
\begin{eqnarray}\label{eq:semiC}
\frac{d}{dt} \tilde{a}_0 &=& -(i\Delta_0+\kappa/3)\tilde{a}_0-i\chi_0\tilde{a}_0^{\ast}\tilde{a}_0^2-2i\frac{\sqrt{2\kappa}}{3}\tilde{a}_1+\frac{\sqrt{2\kappa}}{3}(\sqrt{2}i\alpha_0+\alpha_1)\nonumber\\
\frac{d}{dt}\tilde{a}_1 &=& -(i\Delta_1+\kappa/3)\tilde{a}_1-i\chi_1\tilde{a}_1^{\ast}\tilde{a}_1^2-2i\frac{\sqrt{2\kappa}}{3}\tilde{a}_0-\frac{\sqrt{2\kappa}}{3}(\alpha_0+i\sqrt{2}\alpha_1)\nonumber\\
\frac{d}{dt}\langle A_{out,0}\rangle &=&  \frac{-\sqrt{2\kappa}}{3}\tilde{a}_0+2i\frac{\sqrt{\kappa}}{3}\tilde{a}_1+2i\frac{\sqrt{2}}{3}\alpha_0-\frac13\alpha_1\nonumber\\
\frac{d}{dt}\langle A_{out,1}\rangle &=& \frac{-\sqrt{2\kappa}}{3}\tilde{a}_1+2i\frac{\sqrt{\kappa}}{3}\tilde{a}_0-2i\frac{\sqrt{2}}{3}\alpha_1+\frac13\alpha_0.
\end{eqnarray}

In the main Letter, the symbolic semiclassical equations of motion analogous to Eqs.~\eqref{eq:semiC} were produced by QHDL using the slightly more complex schematic in Fig.~\ref{fig2_supp} (which would take up pages of complex expressions to reproduce here -- symbolic algebra capabilities are still a work in progress), which includes transmission line loss and a general phase shift parameter.  Despite their complexity, when numerical parameters were substituted, the resulting nonlinear, complex equations of motion contained only a small number of terms (see below).  These equations of motion were typically integrated numerically in minutes on a laptop, forming the basis of the simulations presented in the main Letter.   

In the actual model used in the main Letter, the model parameters were $\mu=1/\sqrt{2}$, $\nu=i/\sqrt{2}$, phase delays of 2.65 rad, a loss per TKC-pass of 0.4 dB, and the unitless TKC parameters $\kappa = 1/\sqrt{3}$, $\Delta = 0.69$, and $\chi = -4\kappa^2/3\sqrt{3}$ (normalized such that $\Delta = 1$ and $\alpha_{0,1}$=1 correspond to the critical Kerr detuning and drive amplitudes).  The equations of motion that QHDL produces for these parameters are
\begin{eqnarray}\label{eq:actual_model}
\frac{d}{dt}\tilde{a}_0 &=& 0.256600119639834 i \tilde{a}^\ast_0 \tilde{a}^2_0 - 0.269835722981436 \tilde{a}_0 - 0.944502934755685i \tilde{a}_0 - \nonumber\\
&&0.117914147124703 \tilde{a}_1 - 0.582406649882899 i \tilde{a}_1 - 0.16747234932605 \alpha_0 + \nonumber\\
&&0.557479734216854 i \alpha_0 + 
0.383245128440802 \alpha_1 - 0.0775918723951391i \alpha_1\nonumber\\
\frac{d}{dt} \tilde{a}_1 &=& -0.117914147124703 \tilde{a}_0 - 0.582406649882899i \tilde{a}_0 + 0.256600119639834i \tilde{a}_1^\ast \tilde{a}_1^2 - \nonumber\\
&&0.269835722981436 \tilde{a}_1 - 0.944502934755685i \tilde{a}_1 - 0.383245128440802 \alpha_0 + \nonumber\\
&&0.0775918723951391i \alpha_0 + 
0.16747234932605 \alpha_1 - 0.557479734216854i \alpha_1\nonumber\\
\frac{d}{dt}\langle A_{out,0}\rangle &=& -0.286174530945919 \tilde{a}_0 + 0.236841667739385i \tilde{a}_0 + \nonumber\\
&&0.109731478271128 \tilde{a}_1 + 0.541990458354401i \tilde{a}_1 + 0.155850582048067 \alpha_0 + \nonumber\\
&&0.895420212859944i \alpha_0 - 
0.356649778754219 \alpha_1 + 0.0722073734777465i \alpha_1\nonumber\\
\frac{d}{dt}\langle A_{out,1}\rangle &=& -0.286174530945919 \tilde{a}_1 + 0.236841667739385i \tilde{a}_1 + \nonumber\\
&&0.109731478271128 \tilde{a}_0 + 0.541990458354401i \tilde{a}_0 + 0.155850582048067 \alpha_1 -  \nonumber\\
&&0.895420212859944i \alpha_1 -
0.356649778754219 \alpha_0 - 0.0722073734777465i \alpha_0.
\end{eqnarray}

\subsection{Linearized model}
In this section, we primarily describe the linearized model that was used to support the hypothesis that transmission line delays are the main cause for the discrepancy between the observed and simulated output power oscillations (Fig. 5a \& c in the main Letter).  We thank H. Mabuchi for suggesting the outline of this approach.

\begin{figure}[b!]
\includegraphics[width=1\textwidth]{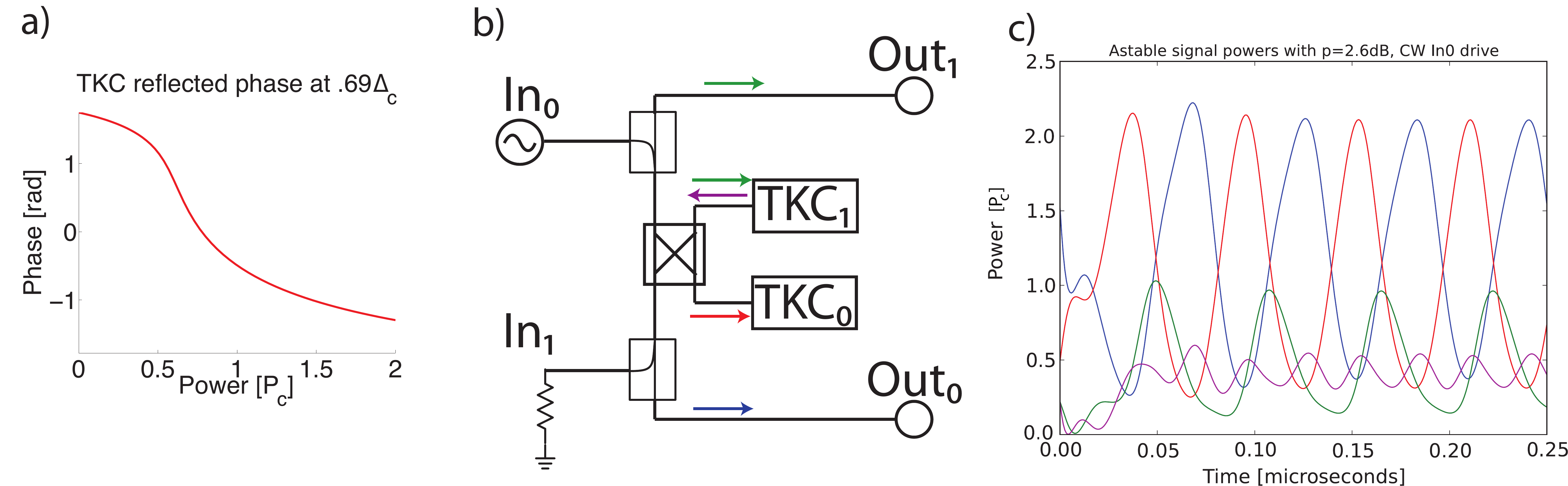}
\caption{\label{fig:astab} a) Steady state reflected phase from a Kerr resonator driven with the experimentally-employed $.69\Delta_c$ detuning.  Note that the x-axis is in units of power, in contrast to Fig.1a in the main Letter.  b) Depiction of the astable configuration presented in the main Letter, with colored arrows indicating various signal segments presented in c.  c) Given an In$_0$ drive of $p=2.58$dB (compare to Fig. 5 in the main Letter) and starting from an un-driven state, a few signal power cycles simulated from the QHDL-derived model are plotted.  As in figure b, red corresponds to the power incident on TKC$_0$, green is both the power exiting Out$_1$ and incident on TKC$_1$, purple is the signal emitted by TKC$_1$, and blue is the signal exiting Out$_0$.}
\vspace{-0.1in}
\end{figure}

\textcolor{black}{First, though, we give a qualitative argument for the delay-induced discrepancy between our system and the model.  For low power In$_0$ drives (and no In$_1$ drive), intra-network signal power is too weak for the TKC non-linearity to be significant.  According to the QHDL-produced model, as the In$_0$ drive increases past $p=1.2$dB, the typical power incident on TKC$_1$ increases past $p=0.4$ (the TKC$_0$ incident power is higher still) and the non-linearity of both resonators becomes significant (see Fig.~\ref{fig:astab}a), leading to the sustained oscillations encountered in the main Letter.  As in the case of bistability, one may roughly understand these astable dynamics through a sequence of events.  As seen in Fig.~\ref{fig:astab}c (where plot colors correspond to the network signals as colored in Fig.~\ref{fig:astab}b), with a sufficiently strong drive on port In$_0$, a rise in the signal power incident on TKC$_0$ (red) results in an increase in both the power exiting Out$_1$ and incident on TKC$_1$ (green) after a characteristic relaxation time.  As the TKC$_1$ incident power rises past $p\approx0.4$, the TKC$_1$ reflected signal (purple) begins to destructively interfere with the drive signal, causing the TKC$_0$ signal to decrease in power and the Out$_0$ signal (blue) to increase.  (Somewhat interestingly, the power of the TKC$_1$ reflected signal is relatively stable, while its phase -- not shown -- varies strongly)  Eventually, as the TKC$_0$ incident power drops, the TKC$_1$ incident power also drops.  The In$_0$ drive then begins to build up the TKC$_0$ incident power again, and the cycle continues.}  

\textcolor{black}{Transmission line delays would allow perturbations to the TKC$_1$ incident power to grow larger before interferometric feedback is able to counteract them, leading to enhanced instability.  For example, in the (no-delay) QHDL model, when the In$_0$ drive is set at $p=-1.6$dB, the steady state TKC$_1$ incident power is $p=0.34$ (not shown).  In the experimental network, delays add up to $0.48\kappa^{-1}=5$ns round trip.  In the sustained oscillations depicted in Fig.~\ref{fig:astab}c, the TKC$_1$ incident power increases from $p=0.34$ to $p=0.7$ in 5ns, well into the non-linear regime for the device.  Thus, one would expect that transmission line delays would lead to stable limit cycles emerging at lower drive powers in general, and that for the experimental system at hand, instability with an In$_0$ drive of only $p=-1.6$dB would not be unreasonable, given typical rates of signal power variation and round trip delays.  These expectations are given a more quantitative foundation in the remainder of this section.}  

The relevant dynamical fixed points of $\{\tilde{a}_0,\tilde{a}_1\}$ for In$_0$ power drives in the range $p = \{-2,5\}$dB were found numerically using Eqs.~\eqref{eq:actual_model}.  We then note that for $\tilde{a}_0 = u_0  + iv_0$, $\tilde{a}_1 = u_1  + iv_1$, the equations of motion for $\{\tilde{a}_0,\tilde{a}_1\}$ from Eqs.~\eqref{eq:actual_model} may be written as
\begin{equation}
\frac{d}{dt}\left[\begin{array}{c} u_0\\v_0\\u_1\\v_1\end{array}\right] = \eta\left[\begin{array}{c} -(u_0^2+v_0^2)v_0\\(u_0^2+v_0^2)u_0\\-(u_1^2+v_1^2)v_1\\(u_1^2+v_1^2)u_1\end{array}\right]+ A^\prime\left[\begin{array}{c} u_0\\v_0\\u_1\\v_1\end{array}\right]+ B\left[\begin{array}{c} \Re\{\alpha_0\}\\\Im\{\alpha_0\}\\\Re\{\alpha_1\}\\\Im\{\alpha_1\}\end{array}\right] 
\end{equation}
where $\eta = 0.256600119639834$, $A^\prime$ and $B$ are $4\times4$ real matrices, $\Re\{\alpha\}$ and $\Im\{\alpha\}$ are the real and imaginary components of $\alpha$, and we have used $i\tilde{a}_i^\ast\tilde{a}_i^2 = i(u_i^2+v_i^2)(u_i+iv_i)$.

The linearized dynamics about the (In$_0$ drive-dependent) fixed points $\{\bar{u}_0,\bar{v}_0,\bar{u}_1,\bar{v}_1\}$ is thus
\begin{eqnarray}\label{eq:int_EOM}
\frac{d}{dt}\left[\begin{array}{c} u_0\\v_0\\u_1\\v_1\end{array}\right] &=& \left(\eta\left[\begin{array}{cccc}
-2\bar{u}_0\bar{v}_0 & -(\bar{u}_0^2+3\bar{v}_0^2) & 0 & 0\\
3\bar{u}_0^2+\bar{v}_0^2 & 2\bar{u}_0\bar{v}_0 & 0 & 0\\
0 & 0 & -2\bar{u}_1\bar{v}_1 & -(\bar{u}_1^2+3\bar{v}_1^2)\\
0 & 0 & 3\bar{u}_1^2+\bar{v}_1^2 & 2\bar{u}_1\bar{v}_1\end{array}\right]+ A^\prime\right)\left[\begin{array}{c} u_0\\v_0\\u_1\\v_1\end{array}\right]+ B\left[\begin{array}{c} \Re\{\alpha_0\}\\\Im\{\alpha_0\}\\\Re\{\alpha_1\}\\\Im\{\alpha_1\}\end{array}\right] \nonumber\\
\frac{d}{dt}\vec{x}&\equiv& \left[\begin{array}{cc} A_{00} & A_{01}\\ A_{10} & A_{11}\end{array}\right]\vec{x} + B\vec{u}
\end{eqnarray}
where we have re-defined the $\{u_i,v_i,\alpha_i\}$ now as deviations about the fixed points, and $\vec{x}$ and $\vec{u}$ are vectors of these deviations.  The $A_{ij}$ are $2\times2$ real matrices, which are dependent on the mean In$_0$ drive through the fixed points.

Similarly, the definition of the output field fluxes $\frac{d}{dt}\langle A_{out,i}\rangle$ from Eqs.~\eqref{eq:actual_model} can be written in matrix form as
\begin{equation}\label{eq:Out_EOM}
\vec{y} = C\vec{x}+D\vec{u}
\end{equation} 
where $\vec{y}=[\frac{d}{dt}\langle A_{out,0}\rangle, \frac{d}{dt}\langle A_{out,1}\rangle]^T$, and $C$ and $D$ are complex $2\times4$ matrices. 

\begin{figure}[b!]
\includegraphics[width=0.75\textwidth]{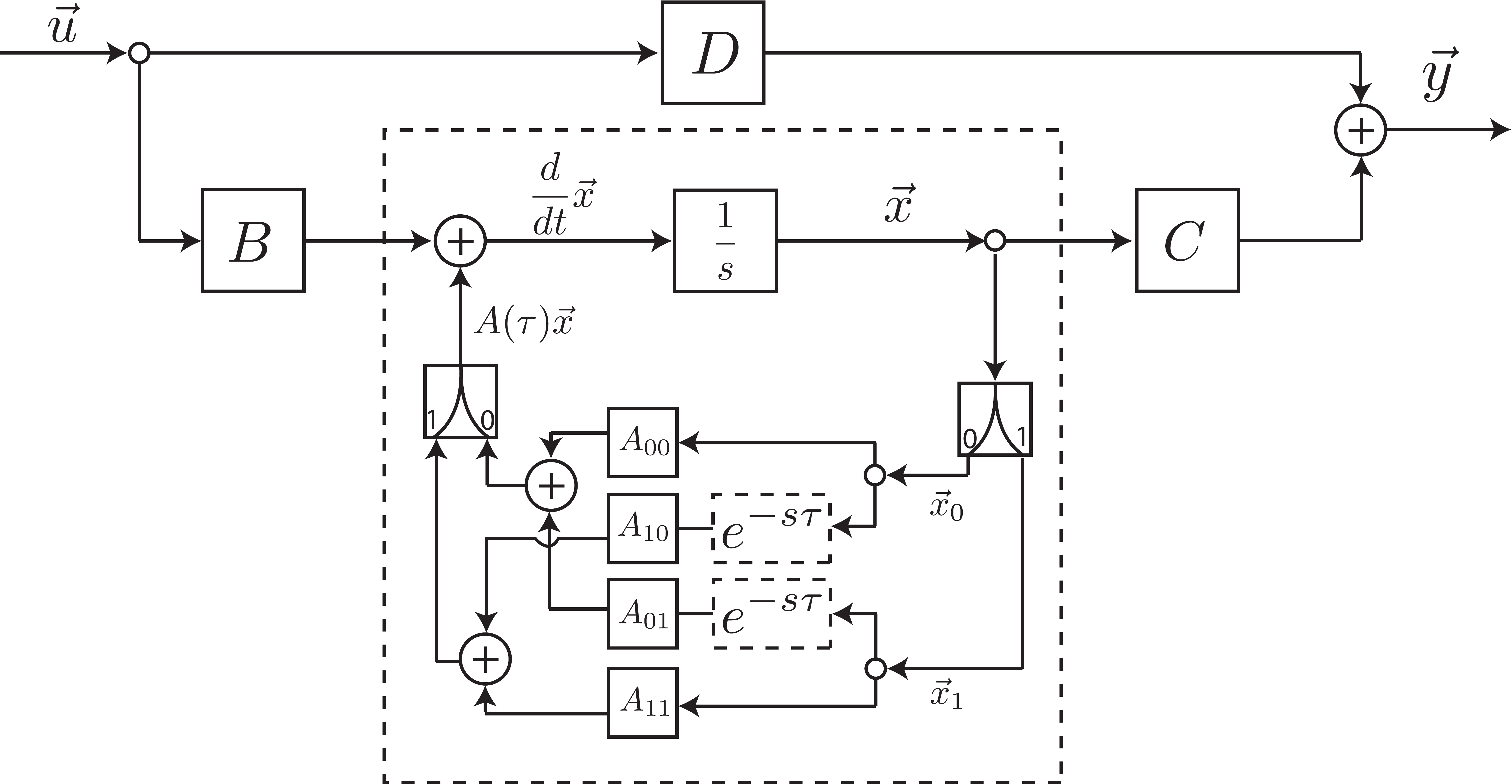}
\caption{\label{fig:FB} Equivalent linearized feedback network suggested by the equations of motion Eqs.~(\ref{eq:int_EOM}-\ref{eq:Out_EOM}).  Using this linear model, the effects of transmission line delays may be approximated by inserting delay `components' $e^{-s\tau}$ in signal lines that represent the driving of the internal state of TKC$_1$ by the internal state of TKC$_0$ and vice versa.}
\vspace{-0.1in}
\end{figure}

We note that the equation of motion Eq.~\eqref{eq:int_EOM} may be modeled as a linear feedback system shown in the dotted box in Fig.~\ref{fig:FB}.  This feedback network represents a system whose input is the B-transformed input deviations, $B\vec{u}$, and whose output is the deviations of the TKCs' internal fields from their fixed points, $\vec{x}$.  Using the linearized network model represented in Fig.~\ref{fig:FB}, we can approximate the consequences of transmission line delays on the overall I/O network dynamics about the calculated fixed points (delays will not effect the fixed point locations).  We make this approximation by inserting delay `components' $e^{-s\tau}$, where $\tau$ is the time delay and $s$ is the Laplace transform variable, on the feedback lines through which the TKC$_0$ deviations, $\vec{x}_0$, drive the TKC$_1$ deviations, $\vec{x}_1$, and vice versa.  This is motivated by the intuition that the dominant contributions to these dynamical `cross terms' have to travel 50 cm of SMA cable (two cable interconnections) in order to drive the dynamics in the other TKC.  Note that even within the linearized model this is an approximation.  For example, the $\vec{x}_0$ contribution that makes multiple `passes' through the network before driving either $\vec{x}_1$ or $\vec{x}_0$ are ignored.  This approximation is justifiable in that the `Q' of the network is very low -- the residual energy left in signals will be low after a few reflections by the 3 dB hybrid.  Using a 5$^{\text{th}}$-order Pad\'{e} transfer function approximation of $e^{-s\tau}$ and the In$_0$ drive-dependent $A_{ij}$ matrices, we can use the Matlab Control Systems Toolbox to calculate a minimal state space model for the feedback network depicted in the dotted box in Fig.~\ref{fig:FB}.  From this model, we can use additional Toolbox functions for I/O pole-zero analysis \cite{Stro01} of the entire linearized network depicted in Fig.~\ref{fig:FB}.  

\begin{figure}[b!]
\includegraphics[width=0.75\textwidth]{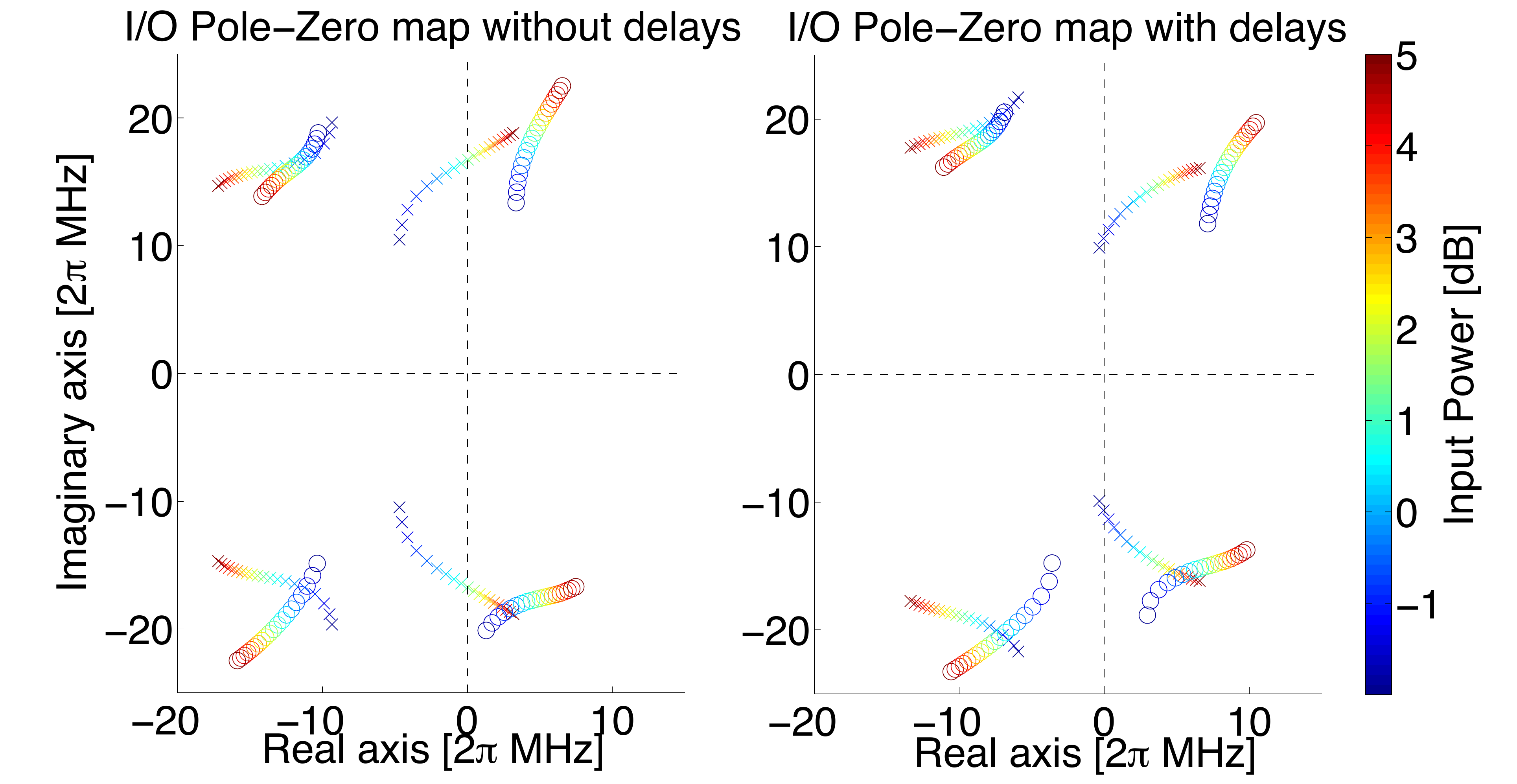}
\caption{\label{fig:PZ_Map} I/O pole-zero maps produced by the Matlab Control Systems Toolbox model of the linear network represented in Fig.~\ref{fig:FB} for various In$_0$ drive powers.  Poles are marked with x's, zeros with o's.  Left, when no transmission line delays are included in the linearized model, a complex pole pair crosses the imaginary axis at $p=$1.2 dB drive power and $\pm2\pi\times17$ MHz, accurately predicting the appearance of the power oscillations observed in model simulation of Fig. 5c in the main Letter.  Right, when 50 cm transmission line delays are included this pole pair is destabilized, crossing the imaginary axis at  -1.6 dB drive power and $\pm2\pi\times11$ MHz, suggesting the appearance of power oscillations much closer to what was observed experimentally in Fig. 5a of the main Letter.}
\vspace{-0.1in}
\end{figure}

After transforming the linearized dynamics back into dimensionfull parameters, in Fig.~\ref{fig:PZ_Map} we plot the pole-zero maps for real In$_0$ drives to Out$_0$ signals for two cases: when transmission line delays are ignored and when they are approximated as described above.  The other I/O maps produce very similar trends.  When no delays are modeled, one sees a marginally stable complex conjugate pole pair move towards the imaginary axis as drive power increases.  At an input drive of $p=$1.2 dB, this pair crosses the imaginary axis with imaginary components $\pm2\pi\times17$ MHz, characteristic of a supercritical Hopf bifurcation that destabilizes the fixed point to a 17 MHz limit cycle.  At higher drives still, the magnitude of the imaginary components of this pair keeps increasing, suggesting that the limit cycle frequency similarly increases.  This interpretation is strongly supported by the simulated power spectrum in Fig. 5c in the main Letter: at 1.2 dB drives, 17 MHz power oscillations suddenly appear and increase in frequency with increasing drive power.  It is well known that the precursors to Hopf bifurcations can be useful for the amplification of AC signals \cite{Jeff85}, suggesting another potential application for our network.  When transmission line delays are approximately modeled, the most conspicuous consequence is to further destabilize this pole pair.  Starting much closer to the imaginary axis, the pair crosses it at -1.6 dB with imaginary components $\pm2\pi\times11$ MHz whose magnitudes increase further with increasing drive power.  This suggests that if transmission line delays were included in the QHDL-produced model, 11 MHz power oscillations would first be observed at -1.6 dB in simulation, much closer to what was experimentally observed in Fig. 5a of the main Letter. 

\end{document}